\documentclass[12pt]{article}
\usepackage[polish,english]{babel}
\usepackage[latin1]{inputenc}
\usepackage{times}
\usepackage[T1]{fontenc}
\usepackage{epsfig}

\begin{document}

\title{FIELD-THEORETIC MODELS WITH    V-SHAPED POTENTIALS \thanks{ Presented by H. A. at the XLV Cracow  School of
Theoretical Physics, Zakopane, Poland, May 28-June 4,  2005}}

\author{H. Arod\'z, $\;\;$ P. Klimas $\;$ and $\;$ T. Tyranowski \\$\;\;$ \\  Institute of Physics,
Jagiellonian University, \\ Reymonta 4, 30-059 Cracow, Poland}

\date{$\;$}

\maketitle

\begin{abstract}
In this lecture we outline main results of our investigations of certain field-theoretic systems which have
V-shaped field potential. After presenting  physical examples of such systems we show that in static problems
the exact ground state value of the field is achieved on a finite distance -- there are no exponential tails.
This applies in particular to  soliton-like object called the topological compacton. Next, we discuss  scaling
invariance which appears when the fields are restricted to small amplitude perturbations of the ground state.
Evolution of such perturbations is governed by a nonlinear equation with a non-smooth term which can not be
linearized even in the limit of very small amplitudes. Finally, we briefly describe self-similar and shock-wave
solutions of that equation.

\end{abstract}

\pagebreak

\section{ Introduction}
 The existence of  ground state, also called the vacuum state, is
one of fundamental features of models in condensed matter physics
as well as in particle physics. The ground states can usually be
found by minimizing an effective potential $V.$ In most cases such
potential is smooth at the minimum. Then, the first derivative of
$V$ at the minimum vanishes, and the square root of the inverse of
the second derivative of $V$ at the minimum yields the basic
(perturbative) length scale in the model.  Such characteristic
length can be finite or infinite, but in either case there are
well-developed, described in numerous textbooks, formalisms for
dealing with such models.  In particular, evolution of small
amplitude perturbations of the ground state is governed by
linearized equations, i.e., by a free field theory which in turn
can be used as the starting point for perturbative treatment of
interactions if the original field equations are nonlinear.

It is easy to point out field-theoretic systems such that  the
standard formalism mentioned above is not applicable: it suffices
to choose the  effective potential which is V-shaped at the
minimum. Then, the left and right first derivatives of the
potential have non-vanishing limits when approaching the vacuum
value of the field, and the second derivative at that point does
not exist. It turns out that there exist models with the V-shaped
field potential which are relatively simple, so that they can be
analyzed in detail.  They have several characteristic features
which seem to be independent of detailed form of the field
potential. We would like to especially emphasize two of them: lack
of exponential tails, and asymptotic scaling symmetry. These facts
make such models quite interesting from theoretical viewpoint.

In this lecture we  outline and summarize main results of our investigations of simple  models with V-shaped
field potentials. The original works have been published in \cite{1, 2}. For completeness of this review we also
mention briefly certain new results \cite{3} which are not published yet.

 We start from the presentation in Section 2 of two macroscopic, mechanical
systems with infinite number of degrees of freedom which lead in a natural way to the V-shaped potentials: 1)
rectilinear set of elastically coupled pendulums which point upwards and bounce from two stiff rods when they
fall down;  2) rectilinear set of elastically coupled balls which vertically bounce from a floor. The first
system exhibits spontaneous symmetry breaking, and it can host static topological defects. Such defects provide
a nice example of the first universal feature of the models with the V-shaped potentials: the lack of
exponential tails in static, finite energy solutions of field equations. This topic is discussed in Section 3.
Equations of motion of the second mechanical system have scaling symmetry which also seems to be universal in
the limit of small amplitudes of perturbations of the ground state. This symmetry and the related self-similar
solutions of pertinent evolution equation are described in Sections 4 and 5. In Section 6 we briefly describe
shock wave solutions which too seem to be generic phenomenon in such models.

\section{ The physical examples }

\subsection{ The system of elastically coupled pendulums  bouncing between two rods}
The first system we consider consists of infinite number of
ordinary pendulums which are attached to a rectilinear wire at the
points $x_i = i a,$ $  i=0, \pm 1, \pm2, ...$. They can swing only
in the plane perpendicular to the wire, see Fig.1.  Here  $a$ is
the constant distance between any two neighboring pendulums. Each
pendulum has a stiff arm of length $R,$ and the mass $m$ at the
free end, see Fig.2. The wire is elastic with respect to torsion
-- it provides the elastic coupling between the pendulums. There
is one degree of freedom per pendulum: the angle $\phi(x_i,t)$
between the vertical direction and the arm. We adopt the
convention that  $\phi(x_i,t)= 0$ corresponds to the vertical
upward position of the $i$-th pendulum. The vertical downward
position is represented by $\phi = \pi$ and $\phi = - \pi.$
Symmetrically on both sides of the wire, parallel to it we put two
stiff bounding rods which restrict the allowed range of the
angles:
\[|\phi_i(x_i,t)| \leq \phi_0, \;\;\;\; \mbox{where}\;\;\;\; \phi_0 < \pi. \]
Moreover, we  assume that when  a pendulum hits (with a nonzero
velocity)  the rod, it elastically bounces back. The gravitational
force is represented by the acceleration $g$ which has the usual
vertical direction.

\begin{figure}[h!]
\begin{center}\includegraphics[height=100pt,width=300pt]{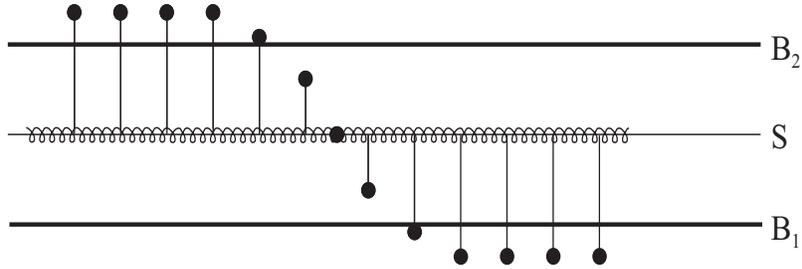}
\end{center}\caption{ The system of pendulums: the view from above. The
central dot represents a pendulum which is directed vertically
upward. Lines $B_1, \: B_2$ represent the two bounding rods, line
$S$ represents the wire to which the pendulums are attached. The
springs on the wire symbolize its torsional elasticity.}
\end{figure}

\floatsep=1.5cm
\begin{figure}[tbh!]
\begin{center}
\includegraphics[height=160pt,width=160pt]{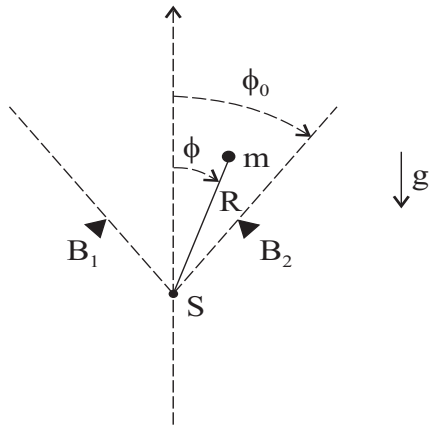}
\end{center}
\caption{ The  perpendicular cross section of the system of
pendulums. Only one pendulum is drawn. $B_1, \: B_2$ and $S$ have
the same meaning as in Fig.1. }
\end{figure}

The system described above reminds the one invented by  A. C.
Scott \cite{4} in order to  demonstrate sinus-Gordon solitons. Our
system has very different properties due to the bounding rods.

When $\phi(x_i,t) < \phi_0$ for all integer $i$ (notice the sharp
inequality), equations of motion for the pendulums have the form
\begin{equation}
\ddot{ \phi}(x_i,t) = \frac{g}{R} \sin \phi(x_i,t) + \kappa
\frac{\phi(x_i- a,t) + \phi (x_i+ a, t) - 2 \phi(x_i,t)}{m a R^2},
\end{equation}
where the dots stand for derivatives  with respect to the time $t$. The first term on the r.h.s. is due to the
gravitational force acting on the mass $m$, and the second one represents the torque of the elastic forces
exerted  by the wire. The constant coefficient $\kappa$ characterizes the torsional elasticity of the wire.

Equation (1) is not valid when $\phi(x_i,t) = \phi_0$ for certain pendulum because it does not include the
instantaneous force acting on that pendulum from the bounding rod. The elastic bouncing condition has the form:
\[
\dot{\phi}(x_i,t) \rightarrow - \dot{\phi}(x_i,t) \] when $\phi(x_i,t) = \pm \phi_0.$

 We will consider Eq. (1) in the continuum limit.
 Standard  steps, see,  the first paper in \cite{1}, yield
\begin{equation}
\frac{\partial^2 \phi(\xi,\tau)}{\partial \tau^2} -
\frac{\partial^2\phi(\xi,\tau)}{\partial \xi^2} - \sin
\phi(\xi,\tau) = 0
\end{equation}
when \[ |\phi(\xi, \tau)| < \phi_0.  \] Here we have introduced the  dimensionless continuous variables:
\[
\tau = \sqrt{\frac{g}{R}}\: t, \;\;\; \xi =
\sqrt{\frac{mgR}{\kappa a}} \:x.\]
 The elastic bouncing condition now takes the form
\begin{equation}  \frac{\partial \phi(\xi, \tau)}{\partial \tau} \rightarrow
- \frac{\partial \phi(\xi, \tau)}{\partial \tau} \;\;\;
\mbox{when} \;\;\; \phi(\xi, \tau) = \pm \phi_0.\end{equation}

Notice that Eq. (2) coincides with the well-known sinus-Gordon equation. In spite of that, it turns out that the
condition (3) and the restriction
\begin{equation}
|\phi(\xi, \tau)| \leq \phi_0
\end{equation}
change the dynamics dramatically.

Equation (2) and  restriction (4) follow from the
  Lagrangian:
\[\;\;\;\; L= \frac{1}{2} (\partial_{\tau}\phi)^2 - \frac{1}{2} (\partial_{\xi}\phi)^2 - V(\phi),
\]
where  \[ V(\phi) = \left\{
\begin{array}{lcl}
\cos\phi -1  &  {\rm for} &  | \phi| \leq \phi_0, \\
\infty  &  {\rm for}  &   |\phi| > \phi_0.
\end{array}
\right.
\]
This potential is shown in  Fig. 3.
\begin{figure}[bth!]
\begin{center}
\includegraphics[height=5cm, width=8cm]{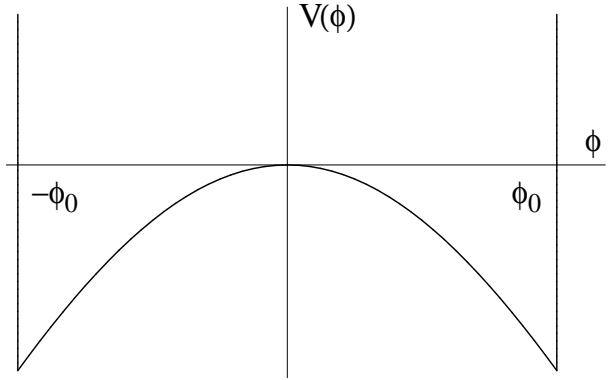}
\end{center}
\caption{ The field potential $V(\phi)$. The physical values of
the field $\phi$ are restricted to the interval $[- \phi_0,
\phi_0].$}
\end{figure}

There are two degenerate ground states: $ \; \phi = \pm \phi_0. $ The Lagrangian has the $Z_2$ symmetry  $ \phi
\rightarrow -\phi$ which is spontaneously broken. Therefore we may expect that in this model  there exists a
topological defect represented by a static solution of field equation (2) interpolating between the two ground
states. Indeed, such defect has been found, \cite{1}. It is described in Section 3 below.

\subsection{The folding transformation}

The bouncing condition (3) implies that  velocities of pendulums
can be discontinuous functions of the time. One can get rid of
this inconvenience by passing to an auxiliary model, which we
shall call the `unfolded' model. It is a new model with a field
$\underline{\phi}(\xi, \tau)$ such that
$\partial_{\tau}\underline{\phi}$ is continuous in $\tau$. As
opposed to $ \phi$, the new field $ \underline{\phi}$ can take
arbitrary real values. Solutions $\phi(\xi, \tau)$ of our original
model are obtained from $\underline{\phi}(\xi, \tau)$  by formulas
which correspond to multiple folding the $\underline{\phi}$ axis,
see Fig. 4. Pertinent formulas can be found in paper \cite{2}.

\begin{figure}[ptbh!]
\begin{center}
\includegraphics[height=250pt,width=400pt]{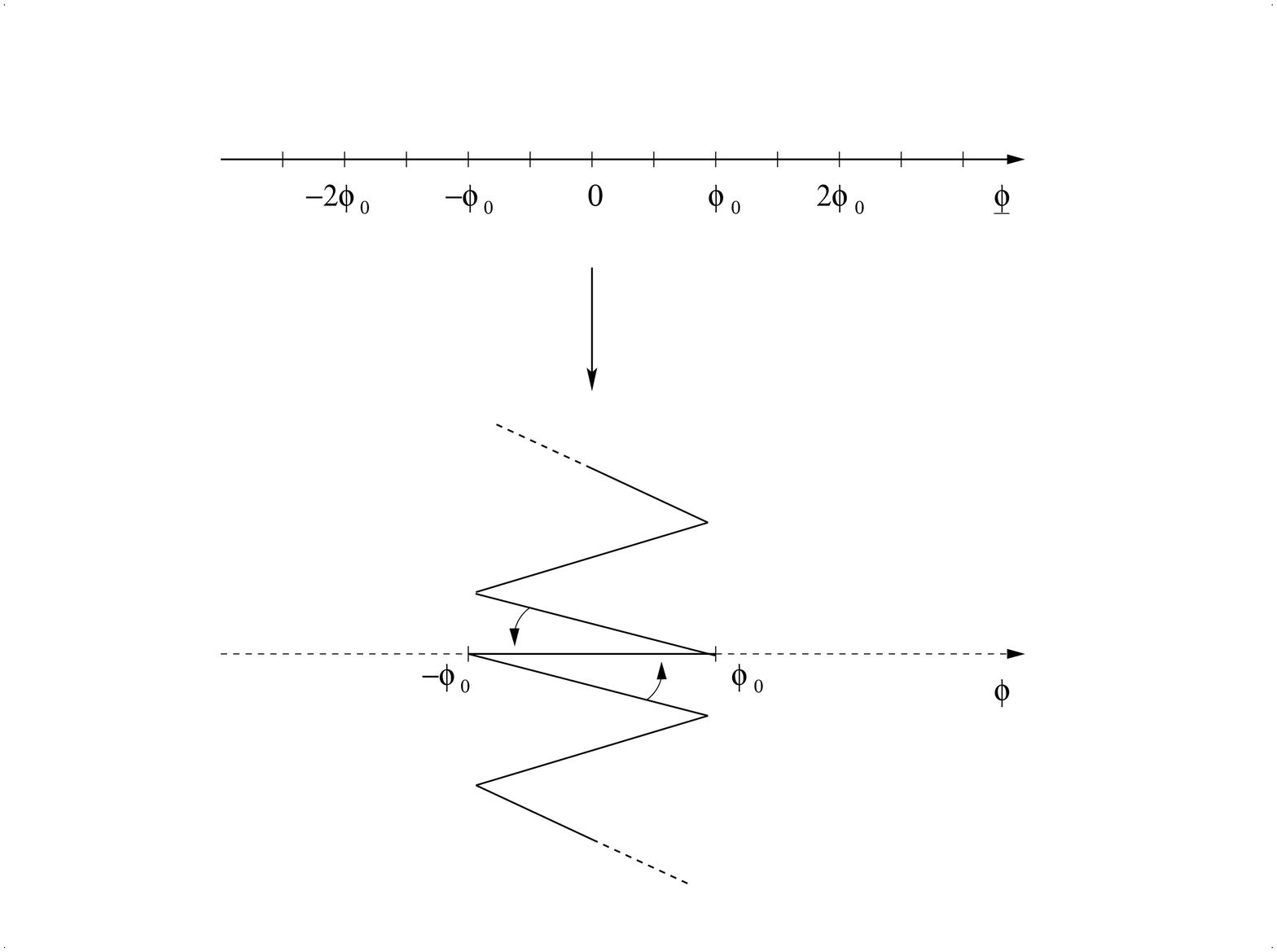}
\end{center}
\caption{  The relation between $\phi$ and $ \underline{\phi}.$}
\end{figure}
\floatsep=0cm
\begin{figure}[ptbh!!]
\begin{center}
\includegraphics[height=180pt,width=330pt]{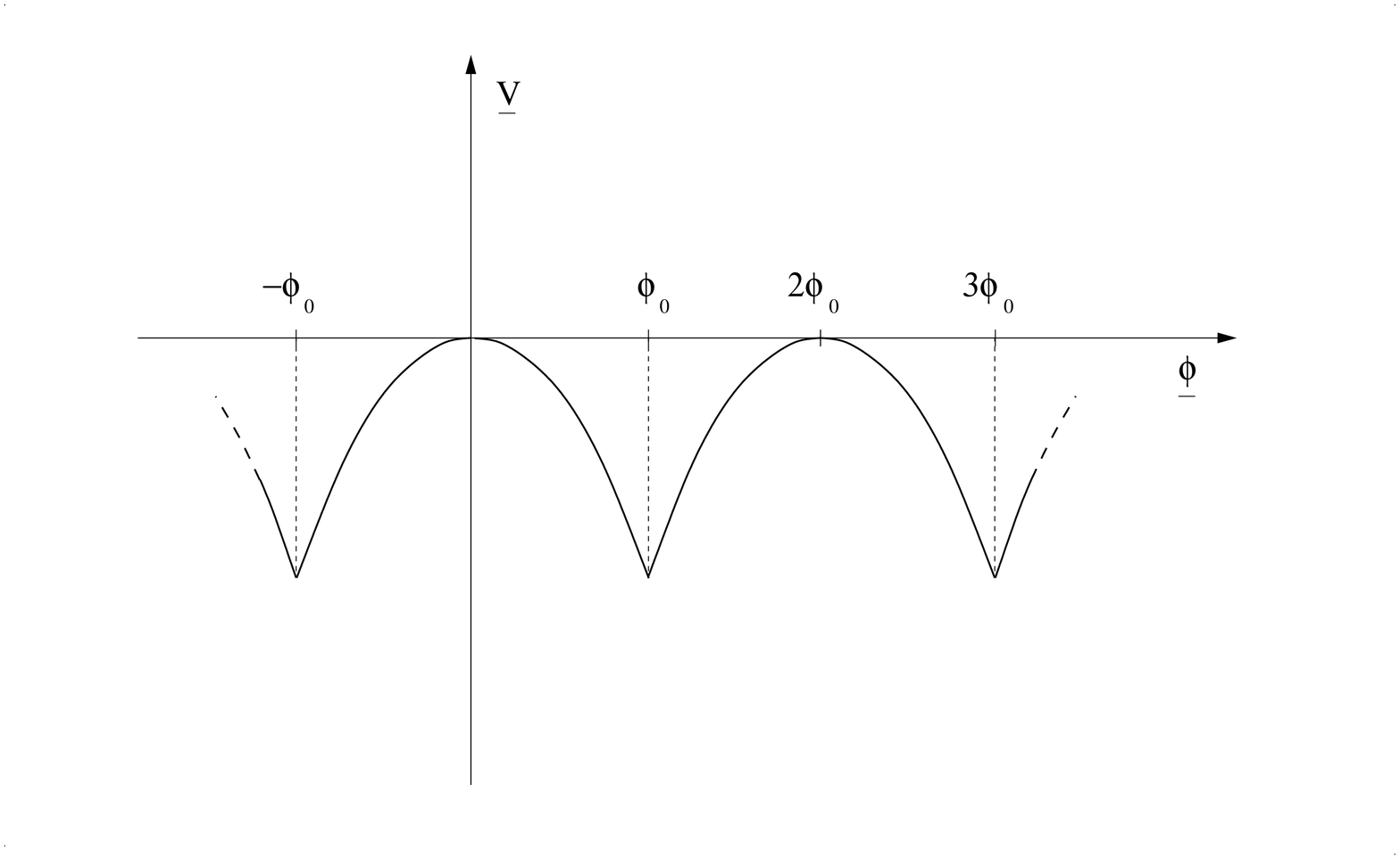}
\end{center}
\caption{ The field potential $\underline{V}(\underline{\phi})$ of
the unfolded model. It is periodic function of
$\underline{\phi}.$}
\end{figure}

 It is clear from Fig. 4 that smooth motion along the $
\underline{\phi}$ axis implies elastic reflection at $\pm \phi_0$
in the  $\phi$ space. The field potential in the unfolded model is
denoted by $\underline{V}(\underline{\phi})$. It has the form
presented in Fig. 5.

 The potential
$\underline{V}(\underline{\phi})$ is symmetrically V-shaped at
each minimum. The force $- d \underline{V}/ d \underline{\phi}$ is
always finite, hence the velocity $\dot{\underline{\phi}}$ does
not have any discontinuities.

Because of its periodicity the potential $\underline{V}$ can be written in the form of Fourier series. Using
such representation one can show that when $\phi_0 \ll 1$ the unfolded model can be regarded as a nonanalytic
perturbation of the sinus-Gordon model, \cite{3}.

The relation between the unfolded model and the original one
reminds  relation between a group and its universal covering.

\subsection{ The system of elastically coupled, vertically bouncing balls}

Balls of mass $m$ can move along vertical poles which are fastened
to a floor  at the points $\xi_i = a i, \;i = 0, \pm1, \pm2,
\ldots,$ lying along a straight line. The balls bounce elastically
from the floor, and each of them is connected with its nearest
neighbors by elastic strings. The system has one degree of freedom
per ball, represented by the elevation $\epsilon$ of the ball
above the floor, see Fig. 6.
\begin{figure}[h]
\begin{center}
\includegraphics[height=200pt,width=330pt]{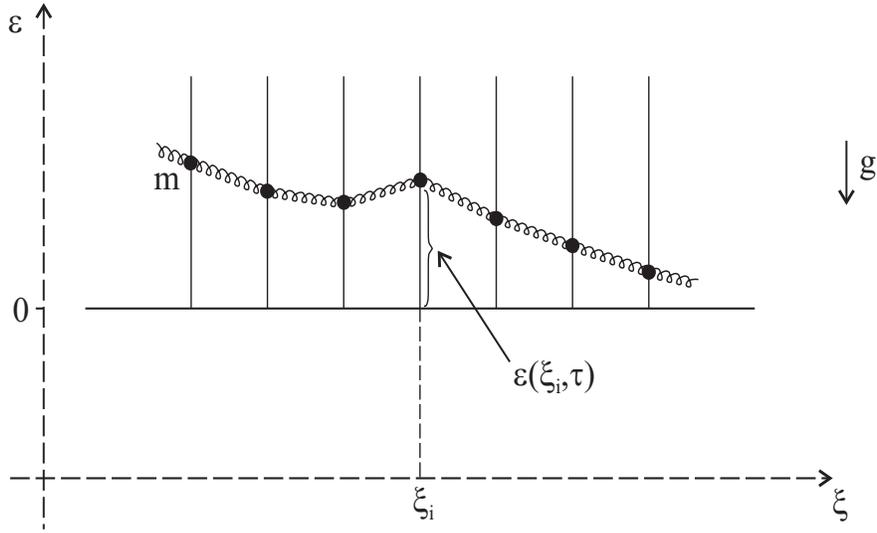}
\end{center}
 \caption{ The  system of balls connected by springs. The balls can move
without friction along the vertical poles (the continuous vertical lines). They elastically bounce from the
floor which is depicted as the continuous horizontal line.}
 \end{figure}

Each ball is subject to the force of gravity and to the elastic
forces from two strings. In a continuum limit equations of motion
for this system can be written in the form
\begin{equation}
\frac{\partial^2 \epsilon(\xi,\tau)}{\partial \tau^2} - \frac{\partial^2\epsilon(\xi,\tau)}{\partial \xi^2} =  -
1,
\end{equation}
where $ \epsilon(\xi, \tau) \geq 0.$

Here we  use dimensionless variables $\tau, \xi$ and $\epsilon$. They differ from the dimensional ones by
appropriate dimensional factors, \cite{3}.

The  elastic bouncing condition has the from:
\[
\frac{\partial \epsilon}{\partial \tau}(\xi, \tau) \rightarrow -  \frac{\partial \epsilon}{\partial
\tau}(\xi,\tau) \;\;\;\; \mbox{when} \;\;\;\; \epsilon(\xi,\tau) = 0. \] Similarly as in the case of pendulums,
we can remove this cumbersome condition by unfolding the model. In the unfolded model, instead of the field
$\epsilon(\xi, \tau) \geq 0$ we have a new field $\underline{\epsilon}(\xi, \tau)$ which can take arbitrary real
values. The evolution equation for $\underline{\epsilon}$ has the form \cite{3}
\begin{equation}
\frac{\partial^2 \underline{\epsilon}(\xi,\tau)}{\partial \tau^2} -
\frac{\partial^2\underline{\epsilon}(\xi,\tau)}{\partial \xi^2} = - \mbox{sign}(\underline{\epsilon}(\xi,\tau)),
\end{equation}
where the $\mbox{sign}$ function has the values $\pm 1$ when $ \underline{\epsilon} \neq 0,$ and 0 if $
\underline{\epsilon} = 0.$  The corresponding field potential has the form
 \begin{equation}
 \underline{V}(\underline{\epsilon}) = | \underline{\epsilon}|.
 \end{equation}
It has the  regular symmetric V shape.
 The folding transformation has the form
 \begin{equation}
 \epsilon(\xi, \tau) = |\underline{\epsilon}(\xi, \tau)|.
 \end{equation}

\section{ The lack of  exponential tails}

Let us assume that the field equation has the following form
\begin{equation}
\frac{\partial^2 \phi(\xi,\tau)}{\partial \tau^2} - \frac{\partial^2\phi(\xi,\tau)}{\partial \xi^2} + V'(\phi) =
0,
\end{equation}
where the field potential $V(\phi)$ is V-shaped at the minimum located at $\phi = \phi_0.$ Hence, in the static
case
\begin{equation}
 \frac{\partial^2\phi(\xi)}{\partial \xi^2}  - V'(\phi) = 0.
\end{equation}
Suppose that $\phi$ approaches the constant ground state value $\phi_0$ from below ($\phi \rightarrow \phi_0-$).
Then Eq. (10) is equivalent to the equation
\[
\frac{\partial \phi(\xi)}{\partial \xi} = \sqrt{ 2 \:( V(\phi) - V(\phi_0))}.
\]
For $\phi$ close to $\phi_0$
\[
\phi(\xi) = \phi_0 - \delta\phi(\xi),
\]
where $ \delta\phi \geq 0,$ and
\[
V(\phi) - V(\phi_0) = - V'(\phi_0-)\: \delta\phi +  \frac{1}{2} V''(\phi_0-)\: (\delta\phi)^2 +  \ldots,
\]
where  $V'(\phi_0-)$ denotes the limit of the first derivative from the side of $\phi <\phi_0$ (the left
derivative of $V$ at $\phi_0$),  $V'(\phi_0-) < 0.$
 Hence, for $ \delta\phi \cong 0$
\[
\partial_{\xi}\delta\phi \cong - \sqrt{2|V'(\phi_0-)|}
\sqrt{\delta\phi}.\] This equation has the general solution of the
form \begin{equation}
 \delta\phi(\xi) \cong
\frac{1}{2} |V'(\phi_0-)| (\xi_0 - \xi)^2,
\end{equation}
where $\xi_0$ is an arbitrary constant.

Thus, we have found a  parabolic approach to the ground state value of the field $\phi$. This value is reached
at $\xi=\xi_0$ exactly -- there is no exponential tail.   The parabolic approach is due to the fact that
$V'(\phi_0-) < 0$. This means that the force pushing  $\phi$ towards  the ground state value does not vanish in
the limit $ \phi \rightarrow \phi_0-.$ In the two physical examples presented above this reflects the fact that
there is a finite threshold for a force which could move the pendulums or the balls upward from the bounding
line or the floor, respectively -- the force has to be stronger than the gravity.

The well-known exponential tails are obtained when $V'(\phi_0)=0$ and $V''(\phi_0) >0$. In that case
\[
\delta\phi(\xi) \cong c_0 \exp(-\sqrt{V''(\phi_0)}\xi ),
\]
where  $c_0$ is a constant.

Nice example of the parabolic approach to the ground state values of the fields is provivded by the topological
defect found in the model with pendulums. Particularly simple is the case when the maximal allowed angle
$\phi_0$ is small, $ \phi_0 \ll 1$. Then $\sin\phi \cong \phi, $ and Eq.(2) may be replaced by the linear one
\begin{equation}
\frac{\partial^2 \phi(\xi,\tau)}{\partial \tau^2} - \frac{\partial^2\phi(\xi,\tau)}{\partial \xi^2} -
\phi(\xi,\tau) = 0.
\end{equation}
However, one should keep in mind that this equation, as well as Eq.(2), is not physically relevant when  $
|\phi(\xi, \tau)| \geq  \phi_0. $ The physically correct configuration is obtained by solving Eq.(12) when $
|\phi| < \phi_0 $ and matching this solution with the ground state fields $ \pm \phi_0.$

Equation (12) has the static solution of the form $ \phi_0 \sin\xi$. Matching it with $\pm \phi_0$ yields the
field $\phi_c(\xi)$ of the topological defect:
 \[ \phi_c(\xi) = \left\{ \begin{array}{ccc} - \phi_0 &\mbox{if} & \xi \leq - \pi/2 \\
\phi_0 \sin\xi & \mbox{if} &  - \pi/2 \leq \xi \leq  \pi/2 \\
+ \phi_0 & \mbox{if} & \xi \geq \pi/2.
\end{array} \right. \]

In the  case $\phi_0$ is not small $\sin\xi$  is replaced by an
elliptic function. Then  the matching with the ground states
fields takes place at $\xi = \pm L/2$, where
\[
L \cong \pi  \left(1 + \frac{\phi_0^2}{16} + \ldots\right) \] when $ \phi_0 \ll 1,$ or
\[
L \cong 2  \ln\frac{4}{\pi - \phi_0}
\] when $ \phi_0 \rightarrow \pi-,$
see \cite{1, 3}.

Energy density for this topological defect reaches the exact
vacuum value at $ \xi = \pm L/2. $ For this reason we call the
defect the `topological compacton'. In literature one can find
other soliton or soliton-like solutions which have compact
support, see, e.g., \cite{5, 6, 7}, but they are not related to
topological sectors in field space. The name `compacton' appears
already in the paper \cite{5}. \\

\begin{figure}[tbph!]
\begin{center} \setlength{\unitlength}{1.0pt}
\begin{picture}(145,100)
 \put(0,50){\vector(1,0){145}} \put(70,0){\vector(0,1){100}} \multiput(70,80)(6,0){8}{\line(1,0){1}}
\multiput(24,20)(6,0){8}{\line(1,0){1}} \put(73,100){$\phi$}
\put(145,52){$\xi$} \put(105,37){$L/2 $}
 \put(40,77){$\phi_0$ = 1} \put(76,17){$-\phi_0$}
\put(15,54){$-L/2$}    \put(120,49){\line(0,1){2}}  \put(23,49){\line(0,1){2}} \put(122,80){\line(1,0){20}}
\put(0,20){\line(1,0){19}}  \multiput(23,21)(0,6){6}{\line(0,1){1}}   \multiput(120,52)(0,6){5}{\line(0,1){1}}
 \qbezier(70,50)(100,80)(124,80)  \qbezier(19,20)(40,20)(70,50)
\end{picture} \end{center}
\caption{ The profile of the topological compacton when $\phi_0
=1$. In this case $L/2 \approx 1.67$.}
\end{figure}
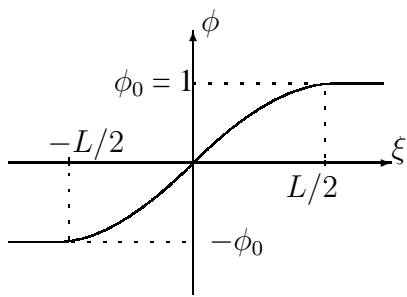

Picture of the compacton in the system of pendulums is provided by
Fig.1. There the compacton is seen from above. At the l.h.s.  of
the picture the pendulums just lie on the bounding rod $B_1$, more
to the center they  gradually rise above $B_1$,  and at the r.h.s.
of the picture they lie on the bounding rod $B_2$.

One  consequence of the lack of  exponential tails is that one can
combine compactons and anti-compactons into a  chain which is
static, because these objects interact only when they touch each
other. The anti-compacton is obtained by reversing the sign of the
field $\phi_c$.

\section{The asymptotic scale invariance}

Let us consider small amplitude oscillations of the field $\phi$ around the V-shaped minimum of the field
potential $V$ at $ \phi = \phi_0:$
\[
\phi(\xi, \tau) = \phi_0 + \psi(\xi, \tau),
\]
where $ |\psi| \ll 1.$ For simplicity we put $\phi_0 =0.$
 Then Eq. (9) acquires the form
\begin{equation}
\frac{\partial^2 \psi(\xi,\tau)}{\partial \tau^2} - \frac{\partial^2\psi(\xi,\tau)}{\partial \xi^2} + V'(\psi) =
0,
\end{equation}
where the field potential $V(\psi)$ is V-shaped at the minimum located at $\psi = 0,$ see Fig. 8. \\
\begin{figure}[tbph!]
\begin{center}
\begin{picture}(200, 110)
\put(0,10){\line(1,0){200}} \put(200,10){\vector(1,0){10}} \put(100,0){\line(0,1){100}}
\put(100,100){\vector(0,1){10}} \put(210,15){\mbox{$\psi$}} \put(105,105){\mbox{$V(\psi)$}}
\put(100,10){\line(2,1){50}} \qbezier(100,10)(160,40)(180,70) \put(100,10){\line(-1,2){40}}
\qbezier(100,10)(60,90)(30,100)
\end{picture}
\end{center}
\caption{Generic V-shaped potential and the piecewise linear
approximation.}
\end{figure}
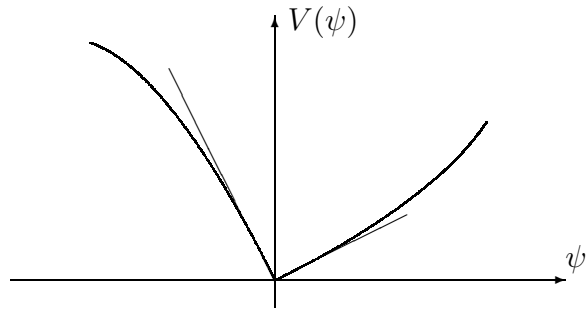

Because  $\psi \approx 0,$ we may replace $V'$ by its piecewise linear approximate  form
\[
V(\psi)\cong V'(0+)\: \psi \: \Theta(\psi) - |V'(0-)|\: \psi \: \Theta(-\psi),
\]
where $\Theta$ denotes the step function.  Then, Eq. (13) is replaced by
\begin{equation}
\frac{\partial^2 \psi }{\partial\tau^2} - \frac{\partial^2
\psi}{\partial\xi^2} = - V'(0+) \:\Theta(\psi) + |V'(0-)|\:
\Theta(-\psi).
\end{equation}
Notice that this equation remains nonlinear for arbitrarily small $\psi$ -- there is no linear regime!

This equation has the scaling symmetry: if $\psi(\xi, \tau)$ is a
solution of it, then
\begin{equation}
\psi_{\lambda}(\xi, \tau) \stackrel{df}{=} \lambda^2 \; \psi(\frac{\xi}{\lambda}, \frac{\tau}{\lambda}),
\end{equation}
where $ \lambda > 0$ is an arbitrary positive number,  obeys Eq. (14)  too. Because this symmetry appears in the
approximate evolution equation (14) we call it the asymptotic scale invariance.

The energy
\[
E[\psi] = \frac{1}{2} \int d\xi \: [ (\partial_{\tau}\psi)^2 +
(\partial_{\xi}\psi)^2 ] + \int d\xi \:V(\psi)
\]
scales as follows:
\[
E[\psi_{\lambda}] = \lambda^3 E[\psi].
\]

In general,  $ \lambda $ should not be  too large because if $
\psi_{\lambda}$ has large amplitude  then  the piecewise linear
approximation is not correct. This restriction does not apply to
the unfolded mechanical model with bouncing balls because in that
case the exact potential (7) is piecewise linear. When $\lambda
\rightarrow 0+ $ the  solutions $\epsilon_{\lambda}(\xi, \tau)$
are in general characterized by high frequencies, short wave
lengths and small energy. This reminds  the phenomenon of
turbulence, even more so when we recall that Navier-Stokes
equations in the high Reynolds number regime have a scale
invariance similar to the one described above.

Immediate consequence of the scale invariance is the lack of a
characteristic frequency for small amplitude oscillations around
the ground state. For example, one can find infinite periodic
running wave solutions to Eq. (2) with the elastic bouncing
condition (3) which have the dispersion relation of the form
\[
\omega^2 - k^2 = \mu^2,
\]
where $\mu^2$ can take any value from the interval $(0, \infty),$
\cite{1, 2}.

\begin{figure}[tbph!]
\begin{center}
\begin{picture}(300, 120)
\put(0,40){\line(1,0){300}} \multiput(150,
22)(0,10){11}{\line(0,1){3}} \put(300,70){\vector(1,0){8}}
\put(150,120){\vector(0,1){8}}
\multiput(0,70)(10,0){30}{\line(1,0){3}}
\put(0,100){\line(1,0){300}} \put(300,75){$\zeta$}   \put(155,
125){$\phi$} \put(155, 105){$\phi_0$}  \put(155, 29){$- \phi_0$}
\qbezier[150](20,100)(40, 60)(60,100) \qbezier[150](60,100)(80,
60)(100,100) \qbezier[150](100,100)(120, 60)(140,100)
\qbezier[150](140,100)(160, 60)(180,100)
\qbezier[150](180,100)(200, 60)(220,100)
\qbezier[150](220,100)(240, 60)(260,100)
\qbezier[30](260,100)(265, 90)(270,85)  \qbezier[30](10,85)(15,
90)(20,100)
\end{picture}
\end{center}

\vspace*{-0.8cm} \caption{ The periodic running wave based on one
bounding rod.}
\end{figure}
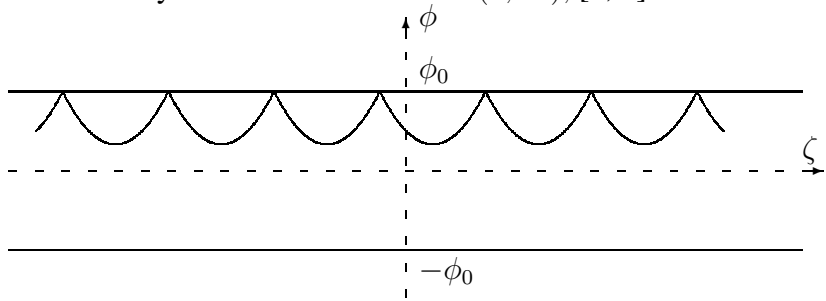
There are two kinds of such waves: with the pendulums bouncing
periodically from one bounding rod, or swinging from one rod to
the other one and back. In the former case the wave has the form
presented in Fig. 9, where
\[  \zeta = \frac{v \tau - \xi}{\sqrt{v^2 -1}}, \]
and  $v >1$ is the phase velocity of the wave.

\section{The  self-similar solutions}

We will discuss self-similar solutions  of Eq. (14) in the
particular case when $V'(0+) = |V'(0-)|.$ After rescaling $\xi$
and $\tau$ by $1/\sqrt{V'(0+)}$ we obtain the following equation:
\begin{equation}
\frac{\partial^2\psi}{\partial\tau^2}  -
\frac{\partial^2\psi}{\partial\xi^2} = - \mbox{sign}(\psi).
\end{equation}
It coincides with Eq.(6) of the unfolded model with the coupled
balls. Direct physical meaning has $|\psi(\xi,\tau)|$ which is
equal to the elevation  above the floor of the ball at the point
$\xi$ at the time $\tau$.

We follow the standard procedure for finding  the self-similar
solutions, see, e.g., \cite{8}. In our case the appropriate Ansatz
for the solution has the form
\[
 \psi(\xi,\tau) = \xi^2 S(y), \;\;\; y = \frac{\tau}{\xi}.
\]
Then Eq.(16) is reduced to the ordinary differential equation for
the function $S(y)$
\[
 (1-y^2)\: S'' +2y\:S' -2\:S = - \mbox{sign}(S).\]
This equation has the following partial solutions:
\[
\mbox{when} \;\;\; S > 0: \;\; \;\;\;\; S(y) =  - \frac{1}{2} \beta (y^2 +1) + \frac{\alpha}{2} y + \frac{1}{2},
\]
\[
\mbox{when} \:\;\; S <0:\;\;\;\;\;\; S(y) =   \frac{1}{2} \beta (y^2 +1) - \frac{\alpha}{2} y - \frac{1}{2},
\]
\[ S(y) = 0, \]
where $\alpha, \beta$ are  constants.  These partial solutions are valid on appropriate finite intervals of the
$y$-axis. They are put together to form the complete solution, \cite{3}. To this end one has to solve recurrence
relations obtained from infinite number of matching conditions. One parameter remains free, so we actually
obtain a family of self-similar solutions. Example of the solution is depicted in Fig. 10.
\begin{figure}
\begin{center}
\begin{picture}(250,80)
\put(0,30){\line(1,0){250}} \put(125,0){\line(0,1){80}}
\put(250,30){\vector(1,0){5}}\put(125,80){\vector(0,1){5}} \put(127, 80){$|\psi|$} \put(253,20){$\xi$}
\put(85,28){\line(0,1){4}} \put(165,28){\line(0,1){4}} \put(74,20){$\xi = - \tau$}  \put(154,20){$\xi = \tau
>0$} \qbezier(170,30)(175,37)(180,30)
\qbezier(180,30)(190,44)(200,30)
\qbezier(200,30)(220,60)(240,30)\qbezier(240,30)(244,35)(248,40)
\qbezier(168,32)(169,31)(170,30) \put(200,50){\vector(1,0){10}}
\put(85,30){\line(1,0){80}} \qbezier(80,30)(81,31)(82,32)
\qbezier(70,30)(75,37)(80,30) \qbezier(50,30)(60,44)(70,30)
\qbezier(10,30)(30,60)(50,30) \qbezier(2,40)(6,35)(10,30)
\put(60,50){\vector(-1,0){10}}
\end{picture}
\end{center}
\caption{ The self-similar solution. The two arrows indicate that
the waves emanate from the points which move along the $\xi$-axis:
$ \xi= \pm \tau$.}
 \end{figure}
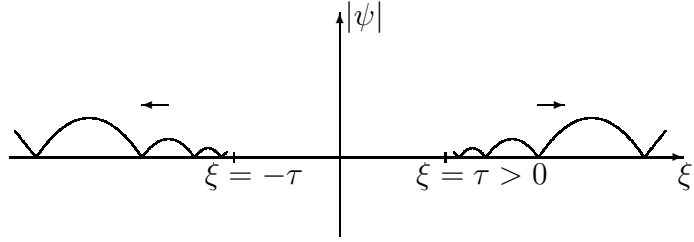
The solution is composed of infinitely many quadratic polynomials
in $\xi$ taken on domains which become smaller and smaller when $
\xi \rightarrow \tau+$ or $ \xi \rightarrow - \tau-$. For  $\xi
\rightarrow \pm \infty$ and fixed $\tau > 0$ the solution
approaches the linear functions $ \pm \tau \xi/2. $ The zeros of
$\psi$ move with `super luminal' velocities, that is with
velocities greater than 1.

The solution presented in Fig. 10 is symmetric with respect to
$\xi \rightarrow - \xi.$ It turns out that the left- and
right-hand halves  of this solution taken separately are solutions
too.

The solution described above cannot be physically realized in the
system of coupled balls because it has infinite energy -- this is
a general feature of self-similar solutions. Nevertheless, finite
pieces of it can be observed. The point is that the dynamics of
the model is local. Therefore, if a physical, finite energy
initial configuration differs from our solution only at very large
values of $|\xi|$,  this difference will be seen in the region of
finite $|\xi|$ only after certain finite time interval. During
that time interval the self-similar solution  describes the
evolution of the part of the system very well.

\section{ The symmetric shock waves}

These solutions are not of the self-similar type.
 The Ansatz
\[
\psi(\xi, \tau) = \Theta(-z) W(z), \;\;\;\;\mbox{where} \;\;\;\; z
= \frac{1}{4} (\xi^2 - \tau^2) \] reduces Eq. (16) to the
following ordinary differential equation
\[ z\:W'' + W' =
\mbox{sign}(W),
\]
which has the following partial solutions:
\[\mbox{when} \;\;\; W > 0: \;\;\;\; W(z) = z + z_1 + z_1
\:\ln|\frac{z}{z_1}| + d_1,
\]
\[
\mbox{when} \;\;\; W < 0: \;\;\;\; W(z) = - z - z_0 - z_0
 \:\ln|\frac{z}{z_0}| - d_0, \]
where $d_0, d_1, z_0, z_1$ are constants. These solutions are
defined on finite intervals of the $z$-axis. Putting them together
 we obtain a one parameter family of solutions which are defined
for all $z\leq 0$, \cite{3}. Because of the step function in the
Ansatz we do not need to know $W(z) $ for $z >0.$

 $\Theta(-z) W(z)$ describes a shock wave, symmetric with respect to $\xi \rightarrow - \xi$ and
restricted to the light-cone $ \xi^2 \leq \tau^2.$ The snapshots
of the wave at three times are shown in Figs. 11, 12, 13.

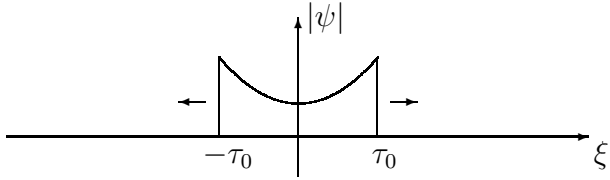
\begin{figure}[tbph!]
\begin{center}
\begin{picture}(240,60)
\put(-30,15){\vector(1,0){220}} \put(80,0){\vector(0,1){60}}
\put(50,15){\line(0,1){30}} \put(110,15){\line(0,1){30}}
\put(45,28){\vector(-1,0){10}}
\qbezier(50,45)(80,10)(110,45)\put(115,28){\vector(1,0){10}}
\put(192,5){$\xi$} \put(44,5){$-\tau_0$} \put(108,5){$\tau_0$}
\put(83,57){$|\psi|$}
\end{picture}
\end{center}
\caption{ The symmetric shock wave at an initial time $\tau_0
>0$.  The arrows indicate that
the wave fronts move. $\psi(\xi, \tau_0) =0 $ for $|\xi| >
\tau_0.$}
\end{figure}
\vspace*{0.5cm}

\begin{figure}[tbph!]
\begin{center}
\begin{picture}(240,60)
\put(-30,15){\vector(1,0){220}} \put(80,0){\vector(0,1){60}}
\put(20,15){\line(0,1){30}} \put(140,15){\line(0,1){30}}
\qbezier(20,45)(30,25)(65,15) \qbezier(65,15)(80,35)(95,15)
\qbezier(95,15)(130,25)(140,45) \put(193,5){$\xi$}
\put(14,5){$-\tau_1$} \put(137,5){$\tau_1$} \put(83,57){$|\psi|$}
\put(15,28){\vector(-1,0){10}} \put(145,28){\vector(1,0){10}}
\end{picture}
\end{center}
\caption{ The symmetric shock wave at a time $\tau_1
> \tau_0.$ Now $\psi(\xi, \tau_1) =0 $ for $|\xi| > \tau_1.$}
\end{figure}
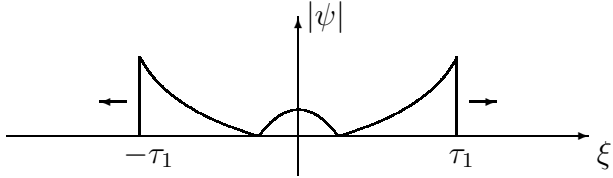
\vspace*{0.5cm}

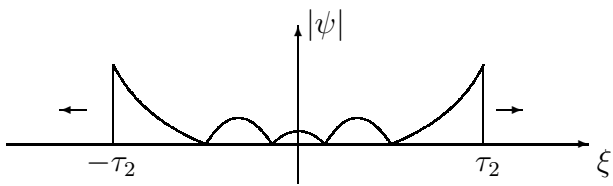
\begin{figure}[tbph!]
\begin{center}
\begin{picture}(240,60)
\put(-30,15){\vector(1,0){220}} \put(80,0){\vector(0,1){60}}
\put(10,15){\line(0,1){30}} \put(150,15){\line(0,1){30}}
\qbezier(10,45)(20,25)(45,15) \qbezier(70,15)(80,25)(90,15)
\qbezier(115,15)(140,25)(150,45) \put(193,5){$\xi$}
\put(0,5){$-\tau_2$} \put(147,5){$\tau_2$} \put(83,57){$|\psi|$}
\put(0,28){\vector(-1,0){10}} \put(155,28){\vector(1,0){10}}
\qbezier(45,15)(57,35)(70,15) \qbezier(90,15)(102,35)(115,15)
\end{picture}
\end{center}
\caption{ The symmetric shock wave at a time $\tau_2 > \tau_1.$ At
later times more zeros of $\psi$ appear.}
\end{figure}
The height of the step, equal to $W(0)$, is the free parameter.
The velocities of the steps (shock fronts) are equal to  $ \pm 1$.
The number of zeros of the function $\psi$ grows indefinitely with
the time $\tau.$ These zeros move along the $\xi$-axis with
velocities  larger that 1, but they never catch up with the front.

The shock wave described above has infinite energy because the
gradient energy density at the shock fronts is infinite. In a real
physical system like the system of coupled balls this energy will
be finite, but then our solution is only an approximation to the
real dynamics. Certainly it can be helpful, but it should be used
with care.

\section{ Summary }

Let us list the main features of classical field-theoretic models
with the V-shaped potentials that we have found:
     \begin{itemize}
     \item{the parabolic approach to the ground state value of the field,}
     \item{the absence of linear regime for oscillations  around the ground state,}
     \item{ the (approximate) scale invariance and the existence of  self-similar solutions,  }
     \item{the existence of shock waves.}
     \end{itemize}
These `signatures' may help to find physical realizations of such
models other than the classical mechanical systems of the type
presented in Section 2.

There are several interesting questions which in our opinion
deserve  attention. \\
1. How does the presence of the threshold force, which is the
reason for the lack of exponential tails,  influence the evolution
of
finite energy perturbations of the ground state? \\
2. The continuum limit yields models which in general are much
simpler than the discrete ones, see, e.g., remarks in \cite{9}.
What are the physical properties of  discrete systems with
V-shaped potentials? In particular,  is there a discrete version
of the shock wave in the system of coupled balls?
\\
3. What are the properties of the corresponding quantum models ?
In particular, one may expect that the scale invariance will be
broken in the quantum model. What is the resulting mass scale? \\

We hope to provide some partial answers soon.

\section{Acknowledgements}
H. Arod\'z and P. Klimas thank the Organizers of the School for
their very kind hospitality, and for the possibility to present
this lecture (H. A.).

This work is supported in part by the COSLAB Programme of the European Science Foundation.

\end{document}